\documentclass[twocolumn,amsmath,amssymb,10pt,aps]{revtex4}
  
\usepackage[utf8]{inputenc}
\usepackage[T1]{fontenc}

\usepackage{amsmath}
\usepackage{amsfonts}
\usepackage{graphicx}
\usepackage{yfonts}
\usepackage{color}
\usepackage[normalem]{ulem}
\usepackage{amsthm}
\usepackage{bm}
\usepackage{bbm}
\usepackage{mathtools}
\usepackage{array}
\usepackage{placeins}
\usepackage{enumitem}

\newcommand{\der}{\mathrm{d}}
\newcommand\bigforall{\mbox{\Large $\mathsurround=1pt\forall$}}

\def\<{\langle}
\def\>{\rangle}

\newcommand{\Tr}{\mathrm{Tr}}
\def\oper{{\mathchoice{\rm 1\mskip-4mu l}{\rm 1\mskip-4mu l}
{\rm 1\mskip-4.5mu l}{\rm 1\mskip-5mu l}}}
\DeclareMathAlphabet\mathbfcal{OMS}{cmsy}{b}{n}

\newtheorem{Theorem}{Theorem}
\newtheorem{Lemma}{Lemma}

\newtheorem{Remark}{Remark}
\newtheorem{Proposition}{Proposition}
\newtheorem{Example}{Example}

\begin{document}

\title{Classical capacity of the generalized Pauli channels}

\author{Katarzyna Siudzi{\'n}ska}
\affiliation{Institute of Physics, Faculty of Physics, Astronomy and Informatics \\  Nicolaus Copernicus University, Grudzi\c{a}dzka 5/7, 87--100 Toru{\'n}, Poland}

%\keywords{quantum channels, quantum information, channel capacity, Holevo capacity, irreducibly covariant channels, Pauli channels, generalized Pauli channels}

\begin{abstract}

We calculate and analyze the bounds of the Holevo capacity and classical capacity for the generalized Pauli channels. In particular, we obtain the lower and upper bounds of the Holevo capacity and show that if these bounds coincide, the Holevo capacity is weakly additive. We find the classical capacity for the Pauli channels and two-parameter generalized Pauli channels.
\end{abstract}

\flushbottom

\maketitle

\thispagestyle{empty}

\section*{Introduction}

In quantum information, one of the fundamental problems is to find the maximal rate of information that can be reliably transmitted by a quantum channel \cite{Nielsen}. This rate is referred to as the {\it channel capacity}, and it is an important quantity considered in the theory of quantum computation and quantum error correction. While the classical channels have a unique definition of capacity, quantum channels can transmit information in a variety of ways. The problem of sending quantum information through a noisy quantum channel was considered by Lloyd \cite{Lloyd}, Shor \cite{Shor}, and Devetak \cite{Devetak2}, who found the lower bound for the quantum capacity. However, if one is interested in transmitting classical information in a quantum state, it is enough to focus on the classical capacity \cite{Holevo,sw}. This is a direct generalization of the classical channel capacity to the quantum scenario. Other communication tasks may require to measure the private classical capacity \cite{Devetak3}, which has its uses in quantum cryptography, or the entanglement-assisted capacity \cite{Bennett}. For more information on the subject, refer e.g. to the review works by Gyongyosi et. al. \cite{Gyongyosi} and Smith \cite{Smith}.

In general, calculating the classical capacity $C(\Lambda)$ of a quantum channel $\Lambda$ is a non-trivial problem. It requires finding an asymptotic limit of the Holevo capacity $\chi(\Lambda)$ for infinitely many uses of the channel. Moreover, the Holevo capacity itself is a maximalization, calculated over all possible ensembles of input quantum states, of the entropic expression given by the Holevo-Schumacher-Westmoreland theorem \cite{Holevo,sw}. The task of obtaining the exact value of the classical capacity simplifies significantly if one considers irreducibly covariant quantum channels. Mathematically speaking, a channel $\Lambda$ is irreducibly covariant with respect to an irreducible unitary representation $U$ of a finite group $G$ if it commutes with the unitary transformation $\mathcal{U}[X]:=U(g)XU^\dagger(g)$ for every $g\in G$. Now, if the channel is irreducibly covariant, then its Holevo capacity is linearly proportional to the minimal output entropy \cite{Holevo2000}. Additionally, as long as the minimal output entropy $S_{\rm min}(\Lambda):=\min_\rho S(\Lambda[\rho])$ is weakly additive, one has $C(\Lambda)=\chi(\Lambda)$ \cite{Holevo2005}. The additivity of $S_{\rm min}(\Lambda)$ was first proved by King for the unital qubit channels \cite{King} and the depolarizing channels \cite{King2}. Covariant quantum channels were first analyzed by Holevo, who considered covariant Markovian semigroups and their generators \cite{Holevo1993,CQME}. Nuwairan \cite{23} introduced the EPOSIC channels, which form a set of extreme points of the irreducibly SU(2)-covariant channels. Jen\u{c}ov\'{a} and Pl\'{a}vala \cite{17} provided the optimality conditions for the covariant quantum channel discrimination.

There exists a method of constructing the channels $\Lambda:\mathcal{B}(\mathcal{H})\to\mathcal{B}(\mathcal{H}^\prime)$ that are irreducibly covariant with respect to a given unitary representation $U$ of a finite group $G$ \cite{MSD}. The aforementioned method works under the conditions that $\mathcal{H}=\mathcal{H}^\prime$ and $U\otimes\overline{U}$ is simply reducible. As a special class of covariant channels that satisfy these requirements, consider the channels covariant with respect to the finite group generated by the Weyl operators. These channels are known as the (discrete) Weyl channels or the Weyl-covariant channels \cite{CQME,Amosov,Holevo2005}. Their properties were analyzed in the work by Datta, Fukuda, and Holevo \cite{DFH}. King et. al. \cite{KingMats} obtained the upper bound for the maximal output 2-norm of the Weyl channels. Later, it was shown that the multiplicativity conjecture of the maximal output 2-norm is satisfied if the maximal bound is reached \cite{DFH}, with more examples given by Fukuda and Gour \cite{Fukuda}. In prime dimensions $d=\dim\mathcal{H}$, imposing additional symmetry constraints on the Weyl channels allows one to construct the generalized Pauli channels \cite{ICQC}. These symmetry constraints are strictly connected with the group theoretical properties of the Weyl-covariant channels. Analogical results were obtained for the multipartite Weyl channels in prime power dimensions \cite{PhD}.

The generalized Pauli channels were first considered by Nathanson and Ruskai \cite{Ruskai} as the {\it Pauli diagonal channels constant on axes}. Their construction is mainly based on the sets of mutually unbiased bases \cite{Wootters}. Due to their unique properties, the generalized Pauli channels found many important uses in quantum information theory. Their applications range between the quantum process tomography \cite{QPT}, optimal parameter estimation \cite{GGPC_1}, and geometrical quantum mechanics \cite{KS2}. In the theory of open quantum systems and non-Markovian dynamics, the evolution of the generalized Pauli channels was analyzed in both the time-local \cite{DCKS,ICQC} and memory kernel approach \cite{DCKS2,KSDC}.

In this paper, we find the bounds of the Holevo capacity and classical capacity for the generalized Pauli channels in power prime dimensions. First, we find the lower and upper bounds of the Holevo capacity by generalizing the results for the Weyl channels, which have been recently obtained in \cite{WCHC,WCHC2}. The exact analytical value of the Holevo capacity is known when the lower and upper bounds coincide. This is also the exact value of the classical capacity, as the lower bound of the Holevo capacity is weakly additive. Examples of the generalized Pauli channels with known classical capacity include the Pauli channels and highly-symmetric two-parameter qudit channels. Proofs to the theorems are included in the appendices.

\section*{Generalized Pauli channels}

Let us consider the most general form of a bistochastic quantum channel \cite{King,Landau}
\begin{equation}\label{Pauli}
\Lambda_P[\rho]=\sum_{\alpha=0}^3 p_\alpha \sigma_\alpha \rho \sigma_\alpha
\end{equation}
with the probability distribution $p_\alpha$ and the Pauli matrices $\sigma_0=\mathbb{I}_2$,
\begin{equation}
\sigma_1=\begin{bmatrix} 0 & 1 \\ 1 & 0 \end{bmatrix},\quad
\sigma_2=\begin{bmatrix} 0 & - i \\  i & 0 \end{bmatrix},\quad
\sigma_3=\begin{bmatrix} 1 & 0 \\ 0 & -1 \end{bmatrix}.
\end{equation}
The channel $\Lambda_P$ is known as the {\it Pauli channel}, and it describes a random unitary evolution of a qubit \cite{Scheel}. Such channels are realized when a unitary evolution is disrupted by errors that arise from classical uncertainties, and the dynamics they provide is also called a {\it mixed-unitary evolution} \cite{TQI} or an {\it evolution under random external fields} \cite{Alicki}. Gregoratti and Werner \cite{Gregoratti} showed that random unitary channels can be corrected by using the classical information obtained by measuring the environment. Audenaert and Scheel \cite{Scheel} provided the necessary and sufficient conditions for a quantim channel to be random unitary.

The eigenvalues $\lambda_\alpha$ of the Pauli channel are real, where the eigenvalue equation is given by
\begin{equation}
\Lambda_P[\sigma_\alpha]=\lambda_\alpha\sigma_\alpha,\qquad \lambda_0=1.
\end{equation}
The relationship between $\lambda_\alpha$ and the probability distribution $p_\alpha$ reads
\begin{equation}
\lambda_\alpha = p_0 + 2p_\alpha - \sum_{\beta=1}^3 p_\beta,\qquad\alpha=1,2,3,
\end{equation}
and the inverse relation is
\begin{equation}
\begin{split}
p_0 &= \frac 14(1 + \lambda_1 + \lambda_2 + \lambda_3),\\
p_\alpha &= \frac 14 \left( 1 + 2\lambda_\alpha - \sum_{\beta=1}^3 \lambda_\beta \right),\quad \alpha=1,2,3.
\end{split}
\end{equation}
Moreover, $\Lambda_P$ is completely positive if and only if its eigenvalues satisfy the Fujiwara-Algoet conditions $|1\pm \lambda_3| \geq |\lambda_1 \pm \lambda_2|$ \cite{Fujiwara,King,Szarek}. Equivalently, they can be rewritten as
\begin{equation}\label{Fuji-2}
-1 \leq \sum_{\beta=1}^{3} \lambda_\beta\leq 1+2\min_{\beta>0}\lambda_\beta.
\end{equation}

An important property of the Pauli channel is related to the fact that the eigenbases $\{\psi_0^{(\alpha)},\psi_1^{(\alpha)}\}$ formed from the eigenvectors of $\{\sigma_1,\sigma_2,\sigma_3\}$ are mutually unbiased. Let us recall that two bases are mutually unbiased if their vectors satisfy the conditions
\begin{equation}
\big\<\psi_k^{(\alpha)}\big|\psi_l^{(\alpha)}\big\>=\delta_{kl},\qquad
\big|\big\<\psi_k^{(\alpha)}\big|\psi_l^{(\beta)}\big\>\big|^2=
\frac 1d
\end{equation}
for $\alpha\neq\beta$. The number of mutually unbiased bases (MUBs) $N(d)$ is bounded from above by $N(d)\leq d+1$ \cite{MAX}. If $d$ is a prime number or a power of a prime ($d=s^r$), then $N(d)=d+1$. In these cases, there are known methods to construct maximal sets of mutually unbiased bases \cite{Wootters,MAX}. For composite dimensions, one can always construct three MUBs \cite{MUB-2}. Moreover, if $d=d_1d_2$, then $N(d)\geq\min\{N(d_1),N(d_2)\}$ \cite{MUB-1}.

In constructing the generalized Pauli channels, we consider the $d$-dimensional Hilbert space $\mathcal{H}$ with the maximal number $N(d)=d+1$ of mutually unbiased bases $\{\psi_0^{(\alpha)},\dots,\psi_{d-1}^{(\alpha)}\}$. Denote the corresponding rank-1 projectors by $P_k^{(\alpha)}:=|\psi_k^{(\alpha)}\>\<\psi_k^{(\alpha)}|$. For $k=1,\ldots,d-1$, construct the unitary operators 
\begin{equation}\label{U}
U_\alpha^k=\sum_{l=0}^{d-1}\omega^{kl}P_l^{(\alpha)},\qquad \omega:=e^{2\pi i/d},
\end{equation}
that form an orthogonal operator basis together with the identity operator $\mathbb{I}_d$. Now, the generalized Pauli channel is defined via \cite{Ruskai,DCKS}
\begin{equation}\label{GPC}
\Lambda_{GP}[\rho]=p_0\rho+\frac{1}{d-1}\sum_{\alpha=1}^{d+1}p_\alpha\sum_{k=1}^{d-1}
U_\alpha^k\rho U_\alpha^{k\dagger},
\end{equation}
where $p_\alpha$ is a probability distribution. For $d=2$, the above reduces to the Pauli channel in eq. (\ref{Pauli}). The eigenvalues of $\Lambda_{GP}$ are $(d-1)$-times degenerated and satisfy
\begin{equation}\label{GPC_eigenvalue_eq}
\Lambda[U_\alpha^k]=\lambda_\alpha U_\alpha^k,\qquad k=1,\ldots,d-1,
\end{equation}
together with $\Lambda[\mathbb{I}_d]=\mathbb{I}_d$. They are related to the probability distribution by
\begin{equation}\label{GPC_eigenvalues}
\lambda_\alpha=\frac{1}{d-1}\left[d(p_0+p_\alpha)-1\right],
\end{equation}
and also
\begin{equation}\label{CCC}
\begin{split}
p_0&=\frac{1}{d^2}\left(1+(d-1)\sum_{\alpha=1}^{d+1}\lambda_\alpha\right),\\
p_\alpha&=\frac{d-1}{d^2}\left(1+d\lambda_\alpha-\sum_{\beta=1}^{d+1} \lambda_\beta\right).
\end{split}
\end{equation}
Finally, the generalized Pauli channel is completely positive if and only if the generalized Fujiwara-Algoet conditions \cite{Fujiwara,Ruskai,Zyczkowski}
\begin{equation}\label{Fuji-d}
-\frac{1}{d-1}\leq\sum_{\beta=1}^{d+1}\lambda_\beta\leq 1+d\min_{\beta>0}\lambda_\beta
\end{equation}
are satisfied.

In the case of prime dimensions $d$, the complete set of mutually unbiased bases can be constructed using the Weyl operators $W_{kl}$. For a fixed orthonormal basis $\{|0\>,|1\>,\ldots,|d-1\>\}$ in $\mathbb{C}^d$, one introduces
\begin{equation}
W_{kl}=\sum_{m=0}^{d-1}\omega^{mk}|m\>\<m+l|.
\end{equation}
They provide the set of unitary operators $\{W_{01},\,W_{10},\,W_{11},\,\ldots,\,W_{1,d-1}\}$
whose eigenbases generate $d+1$ mutually unbiased bases. As $d$ is a prime number, the orthogonal unitary basis $\mathcal{B}=\{W_{kl}\,|\,k,l=0,\ldots,d-1\}$ can be divided into $\mathcal{B}=\{\mathbb{I}_d\} \cup \mathcal{B}_1 \cup \ldots \cup \mathcal{B}_{d+1}$. Every $\mathcal{B}_k$ consists in $d-1$ mutually commuting operators $W_{\alpha k, \alpha l}$ with $\alpha=1,\ldots,d-1$ \cite{MAX}. The correspondence between $W_{mn}$ and $U_\alpha^k$ defined in eq. (\ref{U}) is as follows,
\begin{equation}
U_\alpha^k=\omega^{k(k-1)(\alpha-1)/2}W_{k,k(\alpha-1)},\qquad U_{d+1}^k=W_{0k}.
\end{equation}
Hence, for prime dimensions $d$, the generalized Pauli channel is a special case of the Weyl channel \cite{ICQC}
\begin{equation}\label{W}
\Lambda_W[\rho]=\sum_{k,l=0}^{d-1}p_{kl}W_{kl}\rho W_{kl}^\dagger.
\end{equation} 

For dimensions $d=s^r$ with $s$ being a prime number, the complete set of mutually unbiased bases consists in the eigenbases of the tensor products of $s$-dimensional Weyl operators. 
Observe that $\bigotimes_{a=1}^rW_{k_al_a}$ and $\bigotimes_{a=1}^rW_{m_an_a}$ commute if and only if $\sum_{a=1}^rk_an_a=\sum_{a=1}^rm_al_a \ (\mathrm{mod}\, s)$. The general prescription for constructing $N(d)=d+1$ mutually unbiased bases for $d=s^r$ can be found in Ref. \cite{T0,MUB-1,MUB-2}. The simplest case corresponds to $s=r=2$. One finds five sets of mutually commuting tensor products of the Pauli matrices:
\begin{equation}\label{CPP}
\begin{split}
\mathcal{B}_1=\{\sigma_0\otimes\sigma_1,\sigma_1\otimes\sigma_0,\sigma_1\otimes\sigma_1\},\\
\mathcal{B}_2=\{\sigma_0\otimes\sigma_2,\sigma_2\otimes\sigma_0,\sigma_2\otimes\sigma_2\},\\
\mathcal{B}_3=\{\sigma_0\otimes\sigma_3,\sigma_3\otimes\sigma_0,\sigma_3\otimes\sigma_3\},\\ \mathcal{B}_4=\{\sigma_1\otimes\sigma_2,\sigma_2\otimes\sigma_3,\sigma_3\otimes\sigma_1\},\\ \mathcal{B}_5=\{\sigma_2\otimes\sigma_1,\sigma_1\otimes\sigma_3,\sigma_3\otimes\sigma_2\}.
\end{split}
\end{equation}
Hence, the bipartite generalized Pauli channel has the following Kraus representation,
\begin{equation}\label{bipart_Kraus}
\Lambda_{GP}[\rho]=p_0\rho+\frac{1}{3}\sum_{\alpha=1}^5p_\alpha\sum_{k=1}^{3} B_{\alpha,k}\rho B_{\alpha,k},
\end{equation}
where $B_{\alpha,k}$ is the $k$-th element of $\mathcal{B}_\alpha$. Finally, the generalized Pauli channels are a special case of the multipartite Weyl channels
\begin{equation}\label{MW}
\begin{split}
\Lambda_{W}^{(r)}[\rho]
=&\sum_{k_1,l_1,\ldots,k_r,l_r=0}^{s-1}p_{k_1,l_1,\ldots,k_r,l_r}\\&\times
\left(\bigotimes_{a=1}^rW_{k_al_a}\right)\rho\left(\bigotimes_{a=1}^rW_{k_al_a}^\dagger\right).
\end{split}
\end{equation}
If $r=1$, one reproduces $\Lambda_W$ from eq. (\ref{W}).

\section*{Bounds on the Holevo capacity}

The Holevo capacity is a single-use classical capacity of a quantum channel \cite{Holevo,sw}. It is defined as the maximal value of the entropic expression
\begin{equation}
\chi(\Lambda)=\max_{\{p_k,\rho_k\}}\left[S\left(\sum_kp_k\Lambda[\rho_k]\right)
-\sum_kp_kS(\Lambda[\rho_k])\right],
\end{equation}
where the maximum is calculated over the ensembles of separable states $\rho_k$ with the probabilities of occurence $p_k$. In the above formula, $S(\rho):=-\Tr(\rho\ln\rho)$ is the von Neumann entropy.

In general, finding the exact analytical value of $\chi(\Lambda)$ is not an easy task. However, this problem is significantly simplified for unitarily covariant quantum channels, where
\begin{equation}\label{chi}
\chi(\Lambda)=\ln d-\min_\rho S(\Lambda[\rho]).
\end{equation}
Recall that a quantum channel $\Lambda:\mathcal{B}(\mathcal{H})\to\mathcal{B}(\mathcal{H}^\prime)$ is unitarily covariant with respect to the unitary representations $U\in\mathcal{B}(\mathcal{H})$, $V\in\mathcal{B}(\mathcal{H}^\prime)$ of a finite group $G$ if and only if
\begin{equation}
\bigforall_{X\in\mathcal{B}(\mathcal{H})}\bigforall_{g\in G}\quad
\Lambda[U(g)XU^\dagger(g)]=V(g)\Lambda[X]V^\dagger(g).
\end{equation}
We consider a special class of unitarily covariant channels with $\mathcal{H}_1=\mathcal{H}_2\equiv\mathcal{H}$, $\dim\mathcal{H}=d<\infty$, and $V(g)=U(g)$ for all $g\in G$. Let us take $G$ that is the finite group generated by the Weyl operators. Then, $\Lambda$ is the Weyl channel $\Lambda_W$, also known as the {\it Weyl-covariant} channel \cite{CQME,Amosov,Holevo2005}. The method of construction and properties of $\Lambda_W$ related to the group theory were analyzed in Ref. \cite{ICQC}. In particular, it was shown that if a quantum channel is covariant with respect to all $d-1$ unitary representations $U_\alpha(g)=W_{\alpha k,\alpha l}$, $\alpha=1,\ldots,d-1$, of $G$ for prime $d$, then it is the generalized Pauli channel $\Lambda_{GP}$. Similar calculations were repeated for the multipartite Weyl channels with analogical results \cite{PhD}.

Recently, it has been shown in Refs. \cite{WCHC,WCHC2} how to calculate the bounds for the Holevo capacity for the Weyl channels. Using the methods presented therein, we formulate analogical theorems for the generalized Pauli channels.

\begin{Theorem}\label{TH1}
The Holevo capacity $\chi(\Lambda_{GP})$ of the generalized Pauli channel $\Lambda_{GP}$ is bounded from below by
\begin{equation}\label{lower}
\begin{split}
\chi_{\mathrm{low}}(\Lambda_{GP})=\max_{\alpha>0}\Bigg\{&\frac{1+(d-1)\lambda_\alpha}{d}
\ln[1+(d-1)\lambda_\alpha]\\&+\frac{d-1}{d}(1-\lambda_\alpha)\ln(1-\lambda_\alpha)\Bigg\},
\end{split}
\end{equation}
where $\lambda_\alpha$ are the eigenvalues of $\Lambda_{GP}$.
\end{Theorem}

Observe that the quantum evolution given by $\rho^\prime=\Lambda_{GP}[\rho]$ can be equivalently described using 
$d+1$ probability distributions \cite{DCKS2}
\begin{equation}
\pi_k^{(\alpha)}:=\Tr\left(P_k^{(\alpha)}\rho\right).
\end{equation}
The associated probability vectors $\pi^{(\alpha)}=(\pi_0^{(\alpha)},\ldots,\pi_{d-1}^{(\alpha)})^T$ satisfy the classical evolution equations
\begin{equation}
\pi^{\prime(\alpha)}=T^{(\alpha)}\pi^{(\alpha)}\qquad\mathrm{or}\qquad
\pi_k^{\prime(\alpha)}=\sum_{l=0}^{d-1}T_{kl}^{(\alpha)}\pi_l^{(\alpha)}
\end{equation}
with the bistochastic map
\begin{equation}\label{T}
T_{kl}^{(\alpha)}:=\Tr\left(P_k^{(\alpha)}\Lambda_{GP}[P_l^{(\alpha)}]\right)=
\lambda_\alpha\delta_{kl}+\frac 1d (1-\lambda_\alpha).
\end{equation}
Note that $T^{(\alpha)}$ is a classical symmetric channel, and therefore its capacity reads \cite{Cover,WCHC}
\begin{equation}
C(T^{(\alpha)})=\ln d-H(\mathbf{T}_k^{(\alpha)}),
\end{equation}
where $H(\mathbf{T}_k^{(\alpha)}):=-\sum_{l=0}^{d-1}T_{kl}^{(\alpha)}\ln T_{kl}^{(\alpha)}$ is the Shannon entropy of the $k$-th row of $T^{(\alpha)}$. From eqs. (\ref{LPk}) and (\ref{T}), we see that
\begin{equation}
\Lambda_{GP}[P_k^{(\alpha)}]=T_{kk}^{(\alpha)}P_k^{(\alpha)}+\sum_{m\neq k}T_{km}^{(\alpha)}P_m^{(\alpha)},
\end{equation}
which leads to the conclusion that
$H(\mathbf{T}_k^{(\alpha)})=S(\Lambda_{GP}[P_k^{(\alpha)}])$.
Finally, the lower bound of the Holevo capacity for $\Lambda_{GP}$ is equal to
\begin{equation}
\chi_{\mathrm{low}}(\Lambda_{GP})=\max_\alpha C(T^{(\alpha)}).
\end{equation}

\begin{Remark}\label{Rem1}
The lower bound of the Holevo capacity for the generalized Pauli channels can be equivalently written as
\begin{equation}
\chi_{\mathrm{low}}(\Lambda_{GP})=\ln d-\max_\alpha H(\mathbf{T}_k^{(\alpha)}).
\end{equation}
\end{Remark}

To find the upper bound of $\chi(\Lambda_{GP})$, we need the following lemma.

\begin{Lemma}\label{lemma}
The Holevo capacity $\chi(\Lambda_{W}^{(r)})$ of the multipartite Weyl channel $\Lambda_{W}^{(r)}$ is bounded from above by
\begin{equation}\label{Wupper}
\chi_{\mathrm{up}}(\Lambda_{W}^{(r)})=\ln d-H(\zeta(\mathbf{p})),
\end{equation}
where $\zeta(\mathbf{p})$ is the vector whose subsequent components are the sums without repetition of $d$ greatest numbers from the set $\{p_{k_1l_1\ldots k_rl_r}\,|\,k_a=0,\ldots,s-1;\,a=1,\ldots,r\}$.
\end{Lemma}

As every known generalized Pauli channel $\Lambda_{GP}$ is a special case of the multipartite Weyl channel $\Lambda_{W}^{(r)}$, use Lemma \ref{lemma} to formulate Theorem \ref{TH2}.

\begin{Theorem}\label{TH2}
Assume that the eigenvalues of $\Lambda_{GP}$ are in a non-increasing order, $\lambda_1\geq\lambda_2\geq\ldots\geq\lambda_{d+1}$. Then, the Holevo capacity $\chi(\Lambda_{GP})$ of the generalized Pauli channel $\Lambda_{GP}$ is bounded from above by
\begin{equation}\label{upper}
\chi_{\mathrm{up}}(\Lambda_{GP})=\ln d-H[\zeta(\mathbf{p})].
\end{equation}
The Shannon entropy is given by
\begin{equation}
H[\zeta(\mathbf{p})]=-f_1\ln f_1-\sum_{k=2}^d Z_k\ln Z_k
\end{equation}
for $\sum_{\beta=1}^{d+1}\lambda_\beta\geq\lambda_2$,
\begin{equation}\label{33}
H[\zeta(\mathbf{p})]=-\sum_{k=1}^{m-1}z_k\ln z_k
-F_m\ln F_m-\sum_{k=m+1}^{d}Z_k\ln Z_k
\end{equation}
for $\lambda_{m-1}\geq\sum_{\beta=1}^{d+1}\lambda_\beta\geq\lambda_m$, $m=3,\ldots,d$, and
\begin{equation}
H[\zeta(\mathbf{p})]=-\sum_{k=1}^{d-1}z_k\ln z_k-f_{d+1}\ln f_{d+1}
\end{equation}
for $\lambda_d\geq\sum_{\beta=1}^{d+1}\lambda_\beta$.
The newly introduced variables are defined as
\begin{align*}
&Z_k:=\frac 1d \Bigg[1+(d+1-k)\lambda_k+(k-1)\lambda_{k+1}-\sum_{\beta=1}^{d+1}\lambda_\beta\Bigg],\\
&z_k:=\frac 1d \Bigg[1+(d-k)\lambda_k+k\lambda_{k+1}-\sum_{\beta=1}^{d+1}\lambda_\beta\Bigg],\\
&F_k:=\frac 1d [1+(k-1)\lambda_{k+1}+(d-k)\lambda_k],\\
&f_k:=\frac 1d [1+(d-1)\lambda_k].
\end{align*}
\end{Theorem}

The value of the Holevo capacity for the generalized Pauli channels is restricted by the lower and upper bounds. The lower bound is obtained by calculating the von Neumann entropy of $\Lambda_{GP}$ acting on the projectors onto the mutually unbiased bases. The upper bound is linearly dependent on $S(\Lambda_{GP}[\rho_\ast])$, where $\rho_\ast$ is an optimal state. Note that we do not check whether $\rho_\ast$ exists, so it is possible that $\chi_{\mathrm{up}}(\Lambda_{GP})$ is non-reachable in some cases.

\section*{Classical capacity}

The classical capacity $C(\Lambda)$ of a quantum channel $\Lambda$ measures the optimal rate of classical information transition between the sender and receiver under infinitely many uses of the channel. It is related to the Holevo capacity by the asymptotic expression
\begin{equation}
C(\Lambda)=\lim_{n\to\infty}\frac 1n \chi(\Lambda^{\otimes n}).
\end{equation}
The above formula simplifies significantly if the Holevo capacity is weakly additive; that is, if $\chi(\Lambda\otimes\Lambda)=2\chi(\Lambda)$. Then, one simply has $C(\Lambda)=\chi(\Lambda)$, whereas in general $C(\Lambda)\geq\chi(\Lambda)$ \cite{sw}. Observe that for irreducibly covariant quantum channels the additivity of the Holevo capacity is equivalent to the additivity of the minimal output entropy \cite{Holevo2000,MSD}.

In the previous section, we calculated the lower and upper bounds of the Holevo capacity $\chi(\Lambda_{GP})$ for the generalized Pauli channel $\Lambda_{GP}$. Our knowledge about $\chi(\Lambda_{GP})$ translates into the knowledge about the classical capacity $C(\Lambda_{GP})$. Indeed, the classical capacity is always bounded from below by $C_{\rm{low}}(\Lambda_{GP})=\chi_{\rm{low}}(\Lambda_{GP})$ due to $C(\Lambda)\geq\chi(\Lambda)$.

Now, let us further analyze the properties of the Holevo capacity. Consider the generalized Pauli channel for which $\chi(\Lambda_{GP})=\chi_{\rm{low}}(\Lambda_{GP})$ or $\chi(\Lambda_{GP})=\chi_{\rm{up}}(\Lambda_{GP})$. In this case, the exact value of the classical capacity is known if the respective bound $\chi_{\rm{low/up}}(\Lambda_{GP})$ is weakly additive.

\begin{Proposition}\label{Prop1}
The lower bound $\chi_{\rm{low}}(\Lambda_{GP})$ of the Holevo capacity from Theorem \ref{TH1} is weakly additive.
\end{Proposition}

\begin{Remark}
In the proof to Proposition \ref{Prop1}, it is evident that the lower bound $\chi_{\rm{low}}(\Lambda_W)$ for the Weyl channels is not additive due to $\mathbf{T}_k^{(\alpha)}\neq\mathbf{T}_l^{(\alpha)}$ for $k\neq l$.
\end{Remark}

From Proposition \ref{Prop1}, we see that if $\chi(\Lambda_{GP})=\chi_{\rm{low}}(\Lambda_{GP})$, then $C(\Lambda_{GP})=\chi_{\rm{low}}(\Lambda_{GP})$. In general, an analogical expression for the upper bound does not hold, which can be seen in the following example.

\begin{Example}\label{ex}
As an example of the generalized Pauli channel for which $\chi_{\rm up}(\Lambda_{GP})$ is not weakly additive, consider the Pauli channel $\Lambda_P$ ($d=2$) defined by $p_0=1/4$, $p_1=1/2$, $p_2=1/4$, $p_3=0$, or equivalently by
\begin{equation}
\lambda_1=\frac 12,\qquad \lambda_2=0,\qquad\lambda_3=-\frac 12.
\end{equation}
Eq. (\ref{33}) in Theorem \ref{TH2} allows us to calculate
\begin{equation}
\chi_{\rm up}(\Lambda_{P})=\frac 34 \ln 3-\ln 2.
\end{equation}
Now, let us construct the bipartite channel $\Lambda_P\otimes\Lambda_P$. The corresponding vector $\zeta(\mathbf{p})$ of increasingly ordered $p_\alpha p_\beta$ reads
\begin{equation}
\zeta(\mathbf{p})=\frac{1}{16}(4,2,2,2,2,1,1,1,1,0,0,0,0,0,0,0).
\end{equation}
Using Lemma \ref{lemma}, we see that
\begin{equation}
\chi_{\rm up}(\Lambda_{P}\otimes\Lambda_{P})=\frac{15}{16}\ln 5-\frac{11}{8}\ln 2\neq 2\chi_{\rm up}(\Lambda_{P}).
\end{equation}
\end{Example}

Another corollary from Proposition \ref{Prop1} is that the exact analytical value of the classical capacity $C(\Lambda_{GP})=\chi_{\rm up}(\Lambda_{GP})$ is obtained if the lower and upper bounds of the Holevo capacity coincide. For the generalized Pauli channels, this is the case if their $d+1$ eigenvalues $\lambda_\alpha$ have the same sign and at least $d$ of them have the same values. Namely, if $\lambda_\alpha\leq 0$ and $\lambda_1=\ldots=\lambda_d\equiv\lambda_{\max}$, $\lambda_{d+1}=\lambda_{\min}$, then
\begin{equation}\label{11}
\begin{split}
C(\Lambda_{GP})=&\frac{1+(d-1)\lambda_{\min}}{d}\ln[1+(d-1)\lambda_{\min}]\\&
+(d-1)\frac{1-\lambda_{\min}}{d}\ln[1-\lambda_{\min}].
\end{split}
\end{equation}
Analogically, if $\lambda_\alpha\geq 0$ and $\lambda_1=\lambda_{\max}$, $\lambda_2=\ldots=\lambda_{d+1}\equiv\lambda_{\min}$, then
\begin{equation}\label{22}
\begin{split}
C(\Lambda_{GP})=&\frac{1+(d-1)\lambda_{\max}}{d}\ln[1+(d-1)\lambda_{\max}]\\&
+(d-1)\frac{1-\lambda_{\max}}{d}\ln[1-\lambda_{\max}].
\end{split}
\end{equation}
Note that for $\lambda_1=\ldots=\lambda_{d+1}\equiv\lambda$, eqs. (\ref{22}) and (\ref{22}) recover the classical capacity of the depolarizing channel \cite{King2}.

\subsection*{Special case: Pauli channels}

For $d=2$, the channel capacities have some interesting properties that do not carry over to higher dimensions. First, observe that the lower bound of the Holevo capacity
\begin{equation}\label{lower_Pauli}
\chi_{\rm low}(\Lambda_P)
=\max_{\alpha>0}\left[\frac{1+\lambda_\alpha}{2}\ln(1+\lambda_\alpha)
+\frac{1-\lambda_\alpha}{2}\ln(1-\lambda_\alpha)\right]
\end{equation}
for the Pauli channel $\Lambda_P$ is symmetric with respect to the change of sign $\lambda_\alpha\longmapsto-\lambda_\alpha$. Therefore, the above maximum is reached at $\alpha_\ast$, where $\lambda_{\alpha_\ast}=\max\{|\lambda_{\min}|,\lambda_{\max}\}$. For every Pauli channel, the lower and upper bounds of the Holevo capacity always coincide,
as there exist only two distinct vectors $\zeta(\mathbf{p})$:
\begin{alignat*}{2}
&\zeta(\mathbf{p})=\frac 12 (1+\lambda_{\min},\, 1-\lambda_{\min})\quad&&\mathrm{for}\quad\lambda_{\max}\leq|\lambda_{\min}|,\\
&\zeta(\mathbf{p})=\frac 12 (1+\lambda_{\max},\, 1-\lambda_{\max})\quad&&\mathrm{for}\quad\lambda_{\max}\geq|\lambda_{\min}|.
\end{alignat*}
Hence, the formula for the classical capacity reads
\begin{equation}
C(\Lambda_P)=\frac{1+\lambda_{\alpha_\ast}}{2}\ln(1+\lambda_{\alpha_\ast})
+\frac{1-\lambda_{\alpha_\ast}}{2}\ln(1-\lambda_{\alpha_\ast}),
\end{equation}
where $\lambda_{\alpha_\ast}=\max\{|\lambda_{\min}|,\lambda_{\max}\}$.

\begin{Remark}\label{Rem_fid}
The classical capacity of the Pauli channel is fully determined by its minimal or maximal channel fidelities on pure input states \cite{KS},
\begin{align}
f_{\min}(\Lambda_P)&=\frac 12 \left(1+\lambda_{\min}\right),\\
f_{\max}(\Lambda_P)&=\frac 12 \left(1+\lambda_{\max}\right).
\end{align}
Namely, these quantities are related as follows,
\begin{equation}
C(\Lambda_P)=\ln 2+f_{\ast}\ln f_{\ast}
+(1-f_{\ast})\ln(1-f_{\ast}),
\end{equation}
where
\begin{equation}
f_{\ast}=\begin{cases}
&f_{\min}(\Lambda_P),\qquad\lambda_{\max}\leq|\lambda_{\min}|,\\
&f_{\max}(\Lambda_P),\qquad\lambda_{\max}\geq|\lambda_{\min}|.
\end{cases},
\end{equation}
The channel fidelity is used to measure the distortion of the input states under the action of a given channel. Recently, it has been shown that it can be used to construct a {\it Holevo-like} quantity \cite{Klesse} that is weakly multiplicative.
\end{Remark}

Consider the Pauli dynamical map $\Lambda_P(t)$ that evolves according to the master equation
\begin{equation}\label{ME}
\dot{\Lambda}_P(t)=\mathcal{L}(t)\Lambda_P(t),\quad\Lambda_P(0)=\oper,
\end{equation}
with a time-local generator
\begin{equation}
\mathcal{L}(t)[\rho]=\frac 12 \sum_{\alpha=1}^{3}\gamma_\alpha(t)(\sigma_\alpha\rho\sigma_\alpha-\rho)
\end{equation}
of the Gorini-Kossakowski-Sudarshan-Lindblad form \cite{GKS,L}. If the decoherence rates $\gamma_\alpha(t)$ are non-negative, then the evolution is Markovian \cite{Lt_proof}. The Markovianity of quantum evolution is determined by the divisibility of the associated dynamical map. Namely, the evolution provided by $\Lambda(t)$ is Markovian if and only if $\Lambda(t)=V(t,s)\Lambda(s)$ with a completely positive propagator $V(t,s)$ for any $0\leq s\leq t$. If $V(t,s)$ is positive but not completely positive, then the corresponding $\Lambda(t)$ is P-divisible. For the Pauli channels that solve eq. (\ref{ME}), P-divisibility is equivalent to the lack of information backflow from the environment to the system \cite{Filip}, which is measured by the Breuer-Laine-Piilo distinguishability measure \cite{BLP}. Therefore, $\Lambda_P(t)$ is P-divisible if and only if \cite{ChManiscalco}
\begin{equation}
\frac{\der}{\der t}||\Lambda_P(t)[X]||_1\leq 0
\end{equation}
for any Hermitian operator $X$, where $||X||_1:=\Tr\sqrt{X^\dagger X}$ is the trace norm of $X$. The above condition is equivalent to $\dot{\lambda}_\alpha(t)\leq 0$ \cite{Filip}. Observe that if $\Lambda_P(t)$ is P-divisible, then $\dot{C}[\Lambda_P(t)]\leq 0$, as the classical capacity evolves according to
\begin{equation}
\dot{C}[\Lambda_P(t)]
=\frac{\dot{\lambda}_{\max}(t)}{2}\ln\frac{1+\lambda_{\max}(t)}
{1-\lambda_{\max}(t)},
\end{equation}
where $\lambda_{\max}(t)=\max\{\lambda_\alpha(t)\ |\ \alpha=1,2,3\}$.
The inverse implication is not true. It is necessary for all $\lambda_\alpha(t)$ to be monotonically decreasing in order for the map to be P-divisible, not just for a distinguished $\lambda_{\max}(t)$. This property carries over to the generalized Pauli channels from eq. (\ref{22}).

\section*{Conclusions}

We found the bounds of the Holevo capacity for the generalized Pauli channels, which is the maximal rate of classical information that is reliably transmittable in a single use of a channel. We analyzed these results by showing that, in the most general scenario, the lower bound is weakly additive, contrary to the upper bound. Therefore, whenever both bounds coincide, the analytical value of the classical capacity is known. We presented examples of highly-symmetric generalized Pauli channels, for which it was possible to calculate the classical capacity. Interestingly, the examples included the most general Pauli channels. We showed that if the invertible Pauli dynamical map is P-divisible, then its classical capacity is a monotonously decreasing function of time.

Calculating the exact values of the classical capacity for quantum channel is a very complex task. There are still many open questions that need to be addressed. For one, it would be interesting to find tighter bounds on the Holevo capacity, especially a weakly additive upper bound. Whether the bounds for the Weyl channels or multipartite Weyl channels can be weakly additive requires further studies. We believe that it is possible to obtain the lower bound of the Holevo capacity for the multipartite Weyl channels. First, however, one would have to find the correspondence between the tensor products of the Weyl operators and the projectors onto the mutually unbiased bases. Another open question is whether one can find the classical capacity for more classes of irreducibly covariant quantum channels by using similar methods to the ones presented in this paper.

\section*{Acknowledgements}

This paper was supported by the Polish National Science Centre project No. 2018/28/T/ST2/00008. The author thanks Dariusz Chru{\'s}ci{\'n}ski for valuable discussions.

\bibliography{C:/Users/cynda/OneDrive/Fizyka/bibliography}
\bibliographystyle{beztytulow2}

\section*{Appendix}

\subsection*{Proof to Theorem \ref{TH1}}

The generalized Pauli channels are unitarily covariant with respect to $U_\alpha$, $\alpha=1,\ldots,d+1$. Using eq. (\ref{chi}), one arrives at
\begin{equation}
\begin{split}
\chi(\Lambda_{GP})&=\ln d-\min_\rho S(\Lambda_{GP}[\rho])\\&\geq
\ln d-\min_\alpha S(\Lambda_{GP}[P_k^{(\alpha)}])
=:\chi_{\mathrm{low}}(\Lambda_{GP}).
\end{split}
\end{equation}
It is straightforward to show that
\begin{equation}\label{LPk}
\Lambda_{GP}[P_k^{(\alpha)}]=\frac{1+(d-1)\lambda_\alpha}{d}P_k^{(\alpha)}+\frac{1-\lambda_\alpha}{d}
\sum_{m\neq k}P_m^{(\alpha)},
\end{equation}
from which it follows that the lower bound of the Holevo capacity is given by eq. (\ref{lower}).

\subsection*{Proof to Lemma \ref{lemma}}

Let us briefly recall the proof to Theorem 2 from Ref. \cite{WCHC}. The Holevo capacity of the Weyl channel is given by
\begin{equation}
\chi(\Lambda_W)=\ln d-S(\Lambda_W[\rho_\ast]),
\end{equation}
where $\rho_\ast$ is a pure optimal state minimizing the von Neumann entropy \cite{Wilde}. Assume that the upper bound
\begin{equation}
\chi_{\mathrm{up}}(\Lambda_W)=\ln d-S(\rho_\zeta)
\end{equation}
is achieved at any state with the spectral decomposition
\begin{equation}\label{rhozeta}
\rho_\zeta=\sum_{k=0}^{d-1}q_kQ_k,
\end{equation}
where $Q_k$ are orthogonal rank-1 projectors and $\{q_k\}=\{[\zeta(\mathbf{p})]_k\}$. The condition $\chi(\Lambda_W)\leq\chi_{\mathrm{up}}(\Lambda_W)$ reduces to $S(\rho_q)\leq S(\Lambda[\rho_\ast])$, and finally to
\begin{equation}\label{QQ}
\Lambda[\rho_\ast]\prec \rho_q
\end{equation}
due to the Schur concavity of the von Neumann entropy \cite{Aniello}. The above majorization relation is satisfied if and only if there exists a set of unitary matrices $u_j$ and a probability distribution $s_j$ such that
\begin{equation}
\Lambda_W[\rho_\ast]=\sum_{j=0}^{d-1}s_ju_j\rho_\zeta u_j^\dagger.
\end{equation}
Substituting eq. (\ref{rhozeta}) into the above formula results in the following condition,
\begin{equation}
\sum_{j,k=0}^{d-1}s_jq_ku_jS_k\rho_\ast S_k^\dagger u_j^\dagger=
\sum_{j,k=0}^{d-1}p_{jk}W_{jk}\rho_\ast W_{jk}^\dagger,
\end{equation}
where we used the fact that $Q_k=S_k\rho_\ast S_k^\dagger$, as two pure states differ only by a unitary transformation. Therefore, an admissible choice of $s_j$, $u_j$, and $S_k$ is $s_jq_k=p_{jk}$ and
\begin{equation}\label{us}
u_j=\sum_{m=0}^{d-1}\omega^{jm}|m\>\<m|,\qquad S_k=\sum_{n=0}^{d-1}|n\>\<n+k|.
\end{equation}
Finally, $q_k=\sum_{l=0}^{d-1}p_{jk}$ have to be ordered in a non-increasing way, as $A\prec B$ means that the non-increasingly ordered eigenvalues $\lambda(A)$ of $A$ are majorized by $\lambda(B)$.

The proof for the multipartite Weyl channels is analogical. The main difference is that, instead of the probability distribution $s_j$ and the operators $u_j$, $S_k$ in eq. (\ref{us}), one has $s_{j_1\ldots j_r}q_{k_1\ldots k_r}=p_{j_1k_1\ldots j_rk_r}$,
\begin{align}
u_{i_1\ldots i_r}&=\bigotimes_{a=1}^r\sum_{m_a=0}^{s-1}\omega^{i_am_a}
|m_a\>\<m_a|,\\\
S_{k_1\ldots k_r}&=\bigotimes_{a=1}^r\sum_{n_a=0}^{s-1}|n_a\>\<n_a+k_a|.
\end{align}

\subsection*{Proof to Theorem \ref{TH2}}

According to Lemma \ref{lemma}, the components of $\zeta(\mathbf{p})$ belong to the set
\begin{equation}
J=\left\{p_0,\frac{p_1}{d-1},\ldots,\frac{p_{d+1}}{d-1}\right\},
\end{equation}
where $|J|=d^2$, and every term $p_\alpha/(d-1)$ appears exactly $d-1$ times. Due to eq. (\ref{GPC_eigenvalues}), we see that if $p_\alpha$ with $\alpha=1,\ldots,d+1$ are ordered non-increasingly, then so are $\lambda_\alpha$. Hence, the results depend only on the value of $p_0$. Namely, one has
\begin{equation}
[\zeta(\mathbf{p})]_1=g_1,\quad[\zeta(\mathbf{p})]_k=Q_k
\end{equation}
for $k=2,\ldots,d,$ and $p_0\geq \frac{p_2}{d-1}$;
\begin{equation}
[\zeta(\mathbf{p})]_k=q_k,\quad[\zeta(\mathbf{p})]_m=G_m,\quad[\zeta(\mathbf{p})]_l=Q_l
\end{equation}
for $k=1,\ldots,m-1$, $l=m+1,\ldots,d$, $\frac{p_{m-1}}{d-1}\geq p_0\geq \frac{p_m}{d-1}$, and $m=3,\ldots,d$; and finally
\begin{equation}
[\zeta(\mathbf{p})]_k=G_k,\quad[\zeta(\mathbf{p})]_d=g_{d+1}
\end{equation}
for $k=1,\ldots,d-1$, and $\frac{p_d}{d-1}\geq p_0$.
The newly introduced symbols are defined by
\begin{align*}
&Q_k:=\frac{d+1-k}{d-1}p_k+\frac{k-1}{d-1}p_{k+1},\\
&q_k:=\frac{d-k}{d-1}p_k+\frac{k}{d-1}p_{k+1},\\
&G_k:=\frac{d-k}{d-1}p_k+p_0+\frac{k-1}{d-1}p_{k+1},\\
&g_k:=p_0+p_k.
\end{align*}
Now, the Shannon entropy of the vector $\zeta(\mathbf{p})$ reads
\begin{equation}
H[\zeta(\mathbf{p})]=-\sum_{k=1}^d[\zeta(\mathbf{p})]_k\ln [\zeta(\mathbf{p})]_k.
\end{equation}
Using eq. (\ref{GPC_eigenvalues}), we can express the above formulas in terms of the eigenvalues $\lambda_\alpha$. It is important to note that $p_0\geq p_\alpha/(d-1)$ translates to $\sum_{\beta=1}^{d+1}\lambda_\beta\geq\lambda_\alpha$.

\subsection*{Proof to Proposition \ref{Prop1}}

It is enough to prove that the Shannon entropy of the row of the associated bistochastic map is weakly additive (see Remark \ref{Rem1}). Observe that the evolution $\rho^\prime=(\Lambda_{GP}\otimes\Lambda_{GP})[\rho]$ is equivalently provided by $d+1$ probability distributions
\begin{equation}
\pi_{kl}^{(\alpha)}:=\Tr\left[(P_k^{(\alpha)}\otimes P_l^{(\alpha)})\rho\right].
\end{equation}
Now, the probability vectors obey the classical evolution equation
\begin{equation}
\pi_{kl}^{\prime(\alpha)}=\sum_{i,j=0}^{d-1}T_{kl,ij}^{(\alpha)}\pi_{ij}^{(\alpha)}
\end{equation}
with the bistochastic map
\begin{equation*}
\begin{split}
T_{kl,ij}^{(\alpha)}:&=\Tr\left[(P_k^{(\alpha)}\otimes P_l^{(\alpha)})(\Lambda_{GP}\otimes\Lambda_{GP})[P_i^{(\alpha)}\otimes P_j^{(\alpha)}]\right]\\&=T_{ki}^{(\alpha)}T_{lj}^{(\alpha)}.
\end{split}
\end{equation*}
Recall that $T_{kl}^{(\alpha)}$ is the bistochastic map associated with the evolution $\rho^\prime=\Lambda_{GP}[\rho]$ and defined in eq. (\ref{T}). The Shannon entropy of the $kd+l$-th row of the map $(T_{kl,ij}^{(\alpha)})$ reads
\begin{equation}
H(\mathbf{T}^{(\alpha)}_{kl})=H(\mathbf{T}^{(\alpha)}_k)+H(\mathbf{T}^{(\alpha)}_l).
\end{equation}
Finally, $H(\mathbf{T}^{(\alpha)}_{kl})$ is weakly additive, because for the generalized Pauli channels $H(\mathbf{T}^{(\alpha)}_k)=H(\mathbf{T}^{(\alpha)}_l)$ for any $k,l=0,\ldots,d-1$.

\end{document}